\documentclass[aps,pra,reprint,superscriptaddress]{revtex4-1}

\usepackage{graphicx}

\begin{document}


\title{Frequency-domain model of optical frequency-comb generation in optical resonators with second- and third-order nonlinearities}

\author{Enxu Zhu}
\affiliation{%
College of Science, Hangzhou Dianzi University, Zhejiang 310018, China
}%

\author{Chaoying Zhao}%
\email{Corresponding author: zchy49@163.com}
\affiliation{%
College of Science, Hangzhou Dianzi University, Zhejiang 310018, China
}%
\affiliation{%
State Key Laboratory of Quantum Optics and Quantum Optics Devices, Institute of Opto-Electronics, Shanxi University, Taiyuan 030006, China
}%

\author{Hebin Li}
\affiliation{
Department of Physics, Florida International University, Miami, Florida 33199, USA
}%

\date{\today}

\begin{abstract}
We developed a frequency-domain model describing optical frequency-comb generation in optical resonators with second- and third-order nonlinearities. Compared with time-domain models, our model in principle allows one to express the cavity dispersion accurately, avoiding the dispersion being truncated beyond a certain order. Moreover, the frequency-domain model can readily include frequency dependence of system parameters, such as the linear absorption and the cavity coupling ratio. To demonstrate the validity of our model, we numerically simulated quadratic combs in a singly resonant second-harmonic generation cavity and Kerr combs in a micro-resonator as two examples. The simulated results obtained from the frequency-domain model agree well with those given by previous time-domain models. A system containing both second- and third-order nonlinearities can give rise to many novel physical dynamics. The developed frequency-domain model will contribute to understanding optical frequency-comb generation assisted by multi-order nonlinear processes in various optical resonators.
\end{abstract}


\maketitle

\section{\label{sec:level1}Introduction}

Optical frequency combs (OFCs) constituted by a set of equally spaced discrete spectral lines have a wide range of applications, such as coherent optical communications \cite{Marin2017}, spectroscopy \cite{Dutt2018}, and quantum optics \cite{Wang2020}. Since 2007, the generation of OFCs in a continuous-wave pumped Kerr (third-order) nonlinear resonator \cite{Haye2007} has attracted significant interest due to the advantages of their large free spectral range and compactness compared with traditional OFC sources based on mode-locked lasers. Ultrahigh quality (Q) factor resonators can enhance the intracavity optical field, leading to a strong light-matter interaction \cite{Kipp2004}. In those resonators, the material Kerr nonlinearity gives rise to OFCs by converting continuous-wave pump energy to signal and idle sidebands via the cascading four-wave mixing effect. Such combs are usually termed as Kerr OFCs. After more than ten years of development, the incoherence and high noise of early Kerr OFCs \cite{Herr2012} have been overcome by entering the single soliton state \cite{Herr2014a}, and several techniques have been developed to stabilize the single soliton state \cite{Bras2016a,Joshi2016,Guo2017}.

Recent experiments have shown that OFCs could also be generated in a continuous-wave pumped quadratic nonlinear resonator through phase-matched second-harmonic generation (SHG) \cite{Ricc2015} or optical parametric oscillation (OPO) \cite{Mosca2018}. In contrast to the Kerr nonlinearity, the quadratic nonlinearity is usually relatively stronger, and thus quadratic OFCs may operate with lower pump power. Another advantage of quadratic OFCs is that the parametric process can simultaneously give rise to two combs around the fundamental and second-harmonic frequencies, respectively. Besides those pure quadratic nonlinear resonators, assisted by the SHG, resonators containing both second- and third-order nonlinearities can generate Kerr combs in the normal dispersion regime where the modulation instability is suppressed in a pure Kerr nonlinear resonator \cite{Xue2017}. Very recently, OFCs spanning two octaves have been observed in a deformed silica microtoroid resonator which has a third-order nonlinearity due to the silica medium, and also has a second-order nonlinearity due to the symmetry-breaking \cite{Hao2020}.

It is important to develop theoretical models for understanding the nature of OFC formations in various resonators. There are usually two theoretical approaches to describe Kerr OFCs. The frequency-domain model uses the coupled-mode equations \cite{Chembo2010}. The time-domain model is based on the Lugiato-Lefever equation (LLE) \cite{Coen2013,Chembo2013} which is also known as the driven and damped nonlinear Schr\"{o}dinger equation. For quadratic OFCs, Ricciardi et al. derived three-wave coupled equations to describe the signal and idle as well as fundamental fields in the singly resonant SHG resonator \cite{Ricc2015}. Subsequently, Mosca et al. generalized the three-wave coupled equations to a general expression for any number of interacting fields \cite{Mosca2016}. By assuming that the second-harmonic field varies slowly, Leo et al. solved the second-harmonic field in the Fourier domain, and finally obtained a single mean-field equation in the time domain to describe the combs generation in the singly resonant SHG resonator \cite{Leo2016a}. In Ref.~\cite{Hansson2017}, Hansson et al. discussed the relation of the time-domain mean-field model to the heuristically derived expression. Following a similar technique in Ref.~\cite{Leo2016a}, the model of quadratic OFCs in the singly resonant OPO cavity was derived. Moreover, quadratic OFC formations in doubly resonant SHG and OPO cavities were discussed \cite{Leo2016,Parra2019}. There is also a single envelope equation to model OFC generation in resonators with quadratic and cubic nonlinearities \cite{Hansson2016}.

However, all these models mentioned above are either time-domain models or frequency-domain models that account for only second-order nonlinearity or third-order nonlinearity. Quadratic OFCs assisted by third-order nonlinearity or Kerr OFCs assisted by second-order nonlinearity may present novel physical dynamics \cite{Xue2017,Hao2020}. The time-domain model usually requires the cavity dispersion to be expressed as the high-order derivative of the propagation constant. When the avoided mode crossing effect exists, the cavity dispersion at specific frequency positions is perturbed \cite{Herr2014}, making it difficult to express the cavity dispersion in the time domain. Moreover, when the bandwidth of combs is large, the frequency dependence of parameters such as the linear absorption and the cavity coupling ratio is not negligible.

In this paper, we report a frequency-domain approach to derive coupled equations with a full map in the frequency domain to model OFC generation in optical resonators with both second- and third-order nonlinearities. The model is based on the coupled amplitude equations combined with the resonator effect. The evolution of optical fields can be described by the coupled mean-field equations, considering that the optical field change is small during each roundtrip. The model is validated by the numerical simulations for quadratic OFCs in a singly resonant SHG resonator and Kerr OFCs. The simulation results agree well with previous theoretical models. A frequency-domain approach is more convenient to account for the frequency dependence of parameters in the model. Unlike the time-domain models in which the cavity dispersion must be truncated beyond a certain order, our model in principle allows one to express the cavity dispersion accurately when the resonant frequencies of the cavity are found by an experiment or a Maxwell's equation eigenvalue solver. With increasing interest in the OFC systems containing both second- and third-order nonlinearities, this work will contribute to modeling  and understanding OFC generation involving multi-order nonlinear processes.

\section{\label{sec:level2}Theoretical model}
\begin{figure}[tb]
\includegraphics[width=240pt]{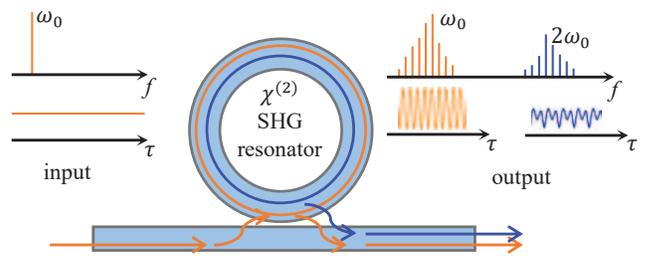}
\caption{\label{fig:1} Schematic of the OFC generation in a ring resonator containing a nonlinear material. A continuous-wave field at frequency $\omega_0$ is injected into the resonator. The quadratic parametric process gives rise to two OFCs around frequency $\omega_0$ and $2\omega_0$ respectively.}
\end{figure}

We consider a nonlinear material in an optical cavity in general. For example, as shown in Fig.~\ref{fig:1}, the nonlinear material is  placed in a ring resonator. The pump field is coupled into the resonator through a waveguide coupler which also serves as an output coupler. The resonator enhances the intracavity field, leading to a stronger light-matter interaction in the cavity. The nonlinear parametric process, such as three-wave mixing for quadratic OFCs and four-wave mixing for Kerr OFCs, converts the input pump power to idler and signal frequency components giving rise to OFCs centered at frequencies $\omega_0$ and $2\omega_0$. In general, the nonlinear material can possess second-, third-, or higher-order nonlinearities.

By expressing polarization $\bf P$ into its linear and nonlinear parts as
\begin{equation}
{\bf P}={\bf P}^{(1)}+{\bf P}^{(2)}+{\bf P}^{(3)}+\cdots={\bf P}^{(1)}+{\bf P}^{\rm NL},
\label{eq:1}
\end{equation}
the optical field in a nonlinear material obeys the nonlinear optical wave equation \cite{Robe2008}
\begin{equation}
\nabla^2{\bf E}-\frac{\epsilon^{(1)}}{c^2}\frac{\partial^2{\bf E}}{\partial t^2}=\frac{1}{\epsilon_0 c^2}\frac{\partial^2 {\bf P}^{\rm NL}}{\partial t^2},
\label{eq:2}
\end{equation}
where $\epsilon^{(1)}$ is the relative permittivity, $c$ is the speed of light in the vacuum, and $\epsilon_0$ is the vacuum permittivity. The relative permittivity $\epsilon^{(1)}$ is given by $\epsilon^{(1)}={\bf [}n+i\alpha_ic/(2\omega){\bf ]}^2$, where $n$ is the linear refractive index of the material, $\alpha_i$ is the linear absorption coefficient, and $\omega$ is the angular frequency of the field.

An optical frequency comb possesses many frequency components and Eq.~(\ref{eq:2}) must hold for each component. In the context of the scalar field approximation, we assume that each component propagates in the z-direction and has the form
\begin{equation}
E_\mu(z,t)={\cal E}_\mu e^{-i\omega_\mu t}+\rm{c.c.},
\label{eq:3}
\end{equation}
where $\mu$ labels various components, c.c. stands for the complex conjugate, and
\begin{equation}
{\cal E}_\mu(z,t)=A_\mu(z) e^{i k_\mu z}.
\label{eq:4}
\end{equation}
Here $A_\mu$ is the amplitude of the field and $k_\mu$ is the magnitude of the wave vector.

For simplicity, here we start by considering only quadratic nonlinearity. The analysis can be extended to the cases that involve higher-order nonlinearities. The nonlinear polarization is represented as
\begin{equation}
{\rm P}_\mu ^{{\rm NL}}={\rm P}_\mu^{(2)}=p_\mu^{(2)}e^{-i\omega_\mu t}+{\rm c.c.}.
\label{eq:5}
\end{equation}
The interaction of all frequency components is considered in
\begin{eqnarray}
p_\mu^{(2)}=&&\epsilon_0\Bigg[\sum_{\alpha,\beta}\chi_{\alpha\beta}^{(2)}{\cal E}_\alpha{\cal E}_\beta\delta\left(\omega_\alpha+\omega_\beta-\omega_\mu\right)\nonumber\\
&&+2\sum_{\gamma,\eta}\chi_{\gamma\eta}^{(2)}{\cal E}_\gamma{\cal E}_\eta^*\delta\left(\omega_\gamma-\omega_\eta-\omega_\mu\right)\Bigg],
\label{eq:6}
\end{eqnarray}
where $\chi^{(2)}$ is the second-order nonlinear optical susceptibility and $\delta (x)$ is the Kronecker delta function. The Kronecker delta function here ensures that only those components satisfying the conservation of energy could contribute to the $\mu$-th component through nonlinear interactions. The first term in Eq.~(\ref{eq:6}) describes the sum-frequency generation process, including the second-harmonic generation as a special case. The second term describes the different-frequency generation which is the inverse process of the first term. There is a factor of 2 in this term since ${\cal E}_\gamma{\cal E}_\eta^*$ and ${\cal E}_\eta{\cal E}_\gamma^*$ are two distinct permutations.

Eqs.~(\ref{eq:3}) to (\ref{eq:6}) are substituted into the nonlinear optical wave equation Eq.~(\ref{eq:2}). Considering $k_\mu=n\left(\omega_\mu\right)\omega_\mu/c$ and adopting the slowly varying amplitude approximation $|d^2A_\mu/dz^2|\ll|k_\mu dA_\mu/dz|$, we obtain the coupled amplitude equation
\begin{equation}
\frac{dA_\mu}{dz}=-\frac{\alpha_{i,\mu}}{2}A_\mu+\frac{i\omega_\mu}{2n\left(\omega_\mu\right)c}{\cal N},
\label{eq:7}
\end{equation}
where ${\cal N}=\sum_{\alpha,\beta}\chi_{\alpha\beta}^{(2)}A_\alpha A_\beta e^{i\Delta k_{\alpha\beta-\mu}z}\delta\left(\omega_\alpha+\omega_\beta-\omega_\mu\right)+2\sum_{\gamma,\eta}\chi_{\gamma\eta}^{(2)}A_\gamma A_\eta^* e^{i\Delta k_{\gamma-\eta-\mu}z}\delta\left(\omega_\gamma-\omega_\eta-\omega_\mu\right)$
is defined to represent the nonlinear term with $\Delta k_{\alpha\beta-\mu}=k_\alpha+k_\beta-k_\mu$ and $\Delta k_{\gamma-\eta-\mu}=k_\gamma-k_\eta-k_\mu$ being the wave-vector mismatches.

The effect of the resonator on the evolution of light fields includes mainly two aspects. First, due to the difference between the resonant frequency of the resonator and the frequency of the optical field, there is an accumulated phase detuning after each round-trip. Considering this detuning, an extra term needs to be added into Eq.~(\ref{eq:7}). It becomes
\begin{equation}
\frac{dA_\mu}{dz}=-\left(\frac{\alpha_{i,\mu}}{2}+i\delta_\mu\right)A_\mu+\frac{i\omega_\mu}{2n\left(\omega_\mu\right)c}{\cal N},
\label{eq:8}
\end{equation}
where $\delta_\mu=\left(\omega_{resonant,\mu}-\omega_\mu\right)t_{{\rm R},\mu}/L$ is the averaged phase detuning of the $\mu$-th frequency component where $\omega_{resonant,\mu}$ represents the resonant frequency closest to $\omega_\mu$, $t_{{\rm R},\mu}$ is the round-trip time of the $\mu$-th  frequency component, and $L$ is the length of the resonator. Second, the intracavity fields are partially transmitted out, while the pump field is partially coupled in the cavity. The intracavity field of the $\mu$-th component $A_\mu^{(m+1)}(0)$ at the beginning of the $(m+1)$th round trip is related to the intracavity field $A_\mu^{(m)}(L)$ at the end of the $m$-th round trip as
\begin{equation}
A_\mu^{(m+1)}(0)=\sqrt{1-\theta_\mu}A_\mu^{(m)}(L)+\sqrt{\theta_\mu}A_{in,\mu},
\label{eq:9}
\end{equation}
where $\theta_\mu$ is the coupling ratio and $A_{in,\mu}$ is the amplitude of the pump field. If the pump field is monochromatic, $A_{in,\mu}$ could be real. If the pump field is polychromatic, $A_{in,\mu}$ should be a complex value to include the phase difference between each pump components.

A full map, known as Ikeda map \cite{Ikeda1979} describing the complete dynamics of OFC generation in a nonlinear resonator, can be formed by combining Eq.~(\ref{eq:8}) with the boundary conditions in Eq.~(\ref{eq:9}). The frequency dependence of parameters can be included in Eqs.~(\ref{eq:8}) and (\ref{eq:9}). The coupled first-order ordinary differential equations in Eq.~(\ref{eq:8}) can be solved by using an ordinary differential equations solver such as the fourth-order Runge-Kutta algorithm. The nonlinear term can be calculated by using the Fast Fourier Transform (FFT) algorithm to significantly speed up the calculation \cite{Hansson2014}.

The definition of OFCs requires that all frequency components are equally spaced. This requirement is automatically fulfilled due to the energy conservation condition of the nonlinear parametric process (the Kronecker delta function in Eq.~(\ref{eq:6})). In practice, only the frequency components close to the resonant frequencies of the resonator have a long photon lifetime in the cavity, so we can assume that the frequency components are spaced by one free spectral range (FSR) of the resonator. Therefore, the frequency components are $\omega_\mu=\omega_0+\mu \omega_{\rm FSR}$ with $\mu=0,\pm1,\pm2,\cdots $, where $\mu=0$ denotes the pump component (if the pump is monochromatic). The detuning $\delta_\mu$ in Eq.~(\ref{eq:8}) implies the cavity dispersion information. One may obtain the resonant frequencies by an experiment or a Maxwell's equation eigenvalue solver, and hence the dispersion can be expressed accurately. There is an alternative way to express the dispersion. This way is similar to what is used in the time-domain models. The resonant frequencies can be expanded in a Taylor series
\begin{equation}
\omega_{resonant,\mu}=\omega_{resonant,0}+D_1\mu+\frac{D_2\mu^2}{2}+\frac{D_3\mu^3}{3}+\cdots,
\label{eq:10}
\end{equation}
where $D_1=\omega_{\rm FSR}=2 \pi/t_{\rm R}$ is the FSR, while $D_2$ and $D_3$ are the second-order dispersion and the third-order dispersion, respectively. If the bandwidth of the comb is not too large and the dispersion is far away from the zero dispersion point (ZDP), the third-order and higher-order dispersion can be neglected. The detuning can thus be expressed as $\delta_\mu=\left(\Delta+D_2\mu^2/2\right)t_{{\rm R},\mu}/L$, where $\Delta=\omega_{resonant,0}-\omega_0$ is defined as the pump detuning.

Finally, if the change of each frequency component after one round-trip is very small (it requires the finesse of the resonator ${\cal F}\gg 1$) \cite{Coen2013}, the intracavity optical fields can be averaged into a set of coupled mean-field equations. To do this, Eqs.~(\ref{eq:8}) and (\ref{eq:9}) are rewritten in a more compact form \cite{Xue2017,Xue2019}
\begin{eqnarray}
\frac{dA_\mu}{dz}=&&-\left(\frac{\alpha_{i,\mu}}{2}+i\delta_\mu\right)A_\mu+\frac{i\omega_\mu}{2n\left(\omega_\mu\right)c}{\cal N}\nonumber\\
&&+\sum_{n=-\infty}^{n=+\infty}\left[\sqrt{\theta_\mu}A_{in,\mu}-\left(1-\sqrt{1-\theta_\mu}\right)A_\mu\right]\nonumber\\
&&\times\delta(z-nL).
\label{eq:11}
\end{eqnarray}
The cavity boundary condition Eq.~(\ref{eq:9}) is represented by the discrete term in Eq.~(\ref{eq:11}). Following the technique introduced in Ref.~\cite{Xue2019}, we can separately consider each term in Eq.~(\ref{eq:11}) to derive the coupled mean-field equations. For example, when only the discrete term is considered, the change of the optical field after one round-trip is given by $\Delta A_\mu=\sqrt{\theta_\mu}A_{in,\mu}-\left(1-\sqrt{1-\theta_\mu}\right)A_\mu$. Replacing the finite difference $\Delta A_\mu$ with $L dA_\mu/dz$, we obtain $dA_\mu/dz=\sqrt{\theta_\mu}A_{in,\mu}/L-\theta_\mu A_\mu/(2L)$. Here the approximation $\sqrt{1-\theta_\mu}\approx 1-\theta_\mu/2$ is used. The coupled mean-field equations are then given by
\begin{equation}
\frac{d A_\mu}{dz}=-\left(\alpha_\mu+i\delta_\mu\right)A_\mu+\frac{i\omega_\mu}{2n\left(\omega_\mu\right)c}{\cal N}+\sqrt{\theta_\mu}A_{in,\mu}/L,
\label{eq:12}
\end{equation}
where $\alpha_\mu=\alpha_{i,\mu}/2+\theta_\mu/(2L)$ is defined to represent total losses of the resonator.

\section{\label{sec:level3}Numerical simulation}
\begin{figure}[b]
\includegraphics[width=240pt]{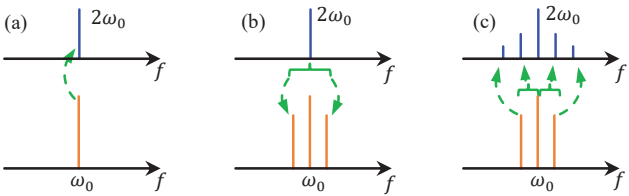}
\caption{\label{fig:2} Schematic of OFC generation via the quasi-phase-matched cavity-enhanced SHG. (a) SHG converts the fundamental pump field to a second-harmonic mode. (b) Once the energy of the second harmonic field reaches the OPO threshold, the second-harmonic modes in turn give rise to an OFC around $\omega_0$. (c) SHG and sum-frequency generation lead to an OFC around $2\omega_0$.}
\end{figure}

In this section, the model is validated by the results from numerical simulations of two examples. The first example is the OFC generation via a quasi-phase-matched cavity-enhanced second-harmonic generation (see Fig.~\ref{fig:1} and Fig.~\ref{fig:2}). The schematic in Fig.~\ref{fig:2} illustrates how a double frequency comb is formed. The second-order nonlinearity first gives rise to a second-harmonic mode. Through the OPO process, the second-harmonic mode is converted back to fundamental components and the fundamental comb is formed. The fundamental comb in turn gives rise to the second-harmonic comb through SHG and sum-frequency generation.

In principle, Eqs.~(\ref{eq:8}) and (\ref{eq:9}) allow us to simulate the comb using the single symbol $A_\mu$. Here we separate the optical fields into fundamental wave and second-harmonic wave to reduce the computational burden. It is valid when there is no additional comb between the fundamental comb and the second-harmonic comb. We use $A_\mu$ to denote the fundamental frequency components  ($A_0$ is corresponding to their center frequency $\omega_0$) and $B_\xi$ to denote the second-harmonic components ($B_0$ is corresponding to their center frequency $2\omega_0$). Considering the quasi-phase-matched condition $\Delta k=0$ and $A_\mu$ and $B_\xi$ are both restricted to their center frequency, Eqs.~(\ref{eq:8}) and (\ref{eq:9}) become
\begin{equation}
\frac{dA_\mu}{dz}=-\left(\frac{\alpha_{iA}}{2}+i\delta_\mu\right)A_\mu+\frac{i\omega_{A,\mu}}{2n_Ac}{\cal N}_A,
\label{eq:13}
\end{equation}
\begin{equation}
\frac{dB_\xi}{dz}=-\left(\frac{\alpha_{iB}}{2}+i\delta_\xi\right)B_\xi+\frac{i\omega_{B,\xi}}{2n_Bc}{\cal N}_B,
\label{eq:14}
\end{equation}
\begin{equation}
A_\mu^{(m+1)}(0)=\sqrt{1-\theta_A}A_\mu^{(m)}(L)+\sqrt{\theta_A}A_{in}\delta(0),
\label{eq:15}
\end{equation}
\begin{equation}
B_\xi^{(m+1)}(0)=\sqrt{1-\theta_B}B_\xi^{(m)}(L),
\label{eq:16}
\end{equation}
where ${\cal N}_A=2\sum_{\gamma,\eta}\chi_Q^{(2)}B_\gamma A_\eta^*\delta(\gamma-\eta-\mu)$ and ${\cal N}_B=\sum_{\alpha,\beta}\chi_Q^{(2)}A_\alpha A_\beta\delta(\alpha+\beta-\xi)$. Here $\chi_Q^{(2)}=(2/\pi)\chi^{(2)}$ is the nonlinear coupling coefficient in the context of the quasi-phase-matched situation \cite{Robe2008}. Moving the pump term in Eq.~(\ref{eq:15}) to Eq.~(\ref{eq:16}) allows one to simulate OFC generation in OPO cavities. To compare with the time-domain model in Ref.~\cite{Leo2016a}, the frequency dependence of parameters in Eqs.~(\ref{eq:13}) to (\ref{eq:16}) are neglected. Using Eq.~(\ref{eq:10}), $\delta_\mu$ and $\delta_\xi$ are given by
\begin{equation}
\delta_\mu=\left(\Delta_A+\frac{1}{2}D_{2A}\mu^2\right)t_{{\rm R},A}/L,
\label{eq:17}
\end{equation}
\begin{equation}
\delta_\xi=\left[\Delta_B+(D_{1B}-D_{1A})\xi+\frac{1}{2}D_{2B}\xi^2\right]t_{{\rm R},B}/L,
\label{eq:18}
\end{equation}
where $\Delta_A=\omega_{resonant,\omega_0}-\omega_0$ and $\Delta_B=\omega_{resonant,2\omega_0}-2\omega_0$. $D_{1A}$ and $D_{1B}$ are the FSR of the fundamental components and second-harmonic components, respectively. $D_{2A}$ and $D_{2B}$ are the second-order dispersion of the fundamental components and second-harmonic components, respectively. It is assumed that the fundamental frequency components are on resonant with the cavity so they are spaced by $D_{1A}$. The second-harmonic components are generated from fundamental components through the parametric process and thus are also spaced by $D_{1A}$ rather than $D_{1B}$. For example, considering two consecutive components generated through $\omega_0+\omega_0\rightarrow 2\omega_0$ and $\omega_0+\left(\omega_0+D_{1A}\right)\rightarrow 2\omega_0+D_{1A}$, the distance between these two second-harmonic components $2\omega_0$ and $2\omega_0+D_{1A}$ is $D_{1A}$.
\begin{figure*}[tb]
\includegraphics[width=380pt]{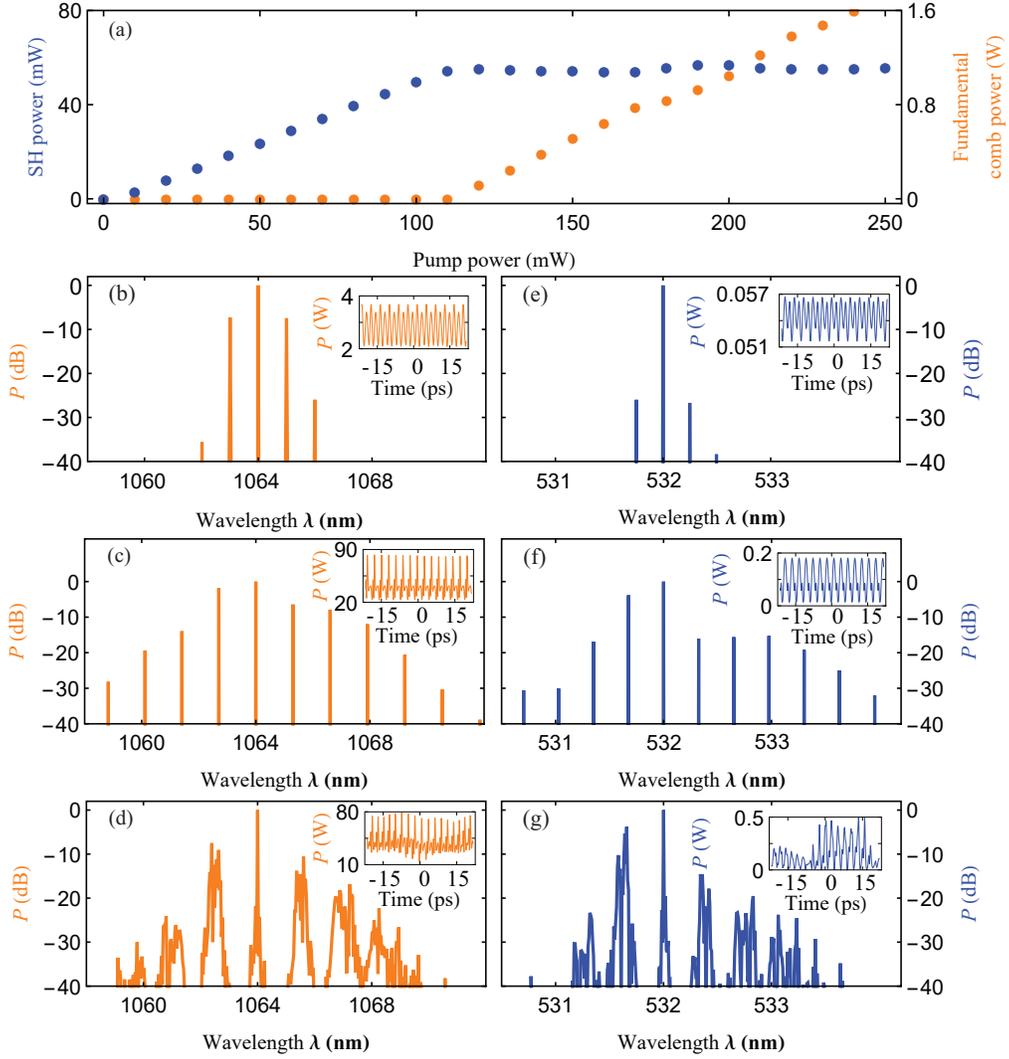}
\caption{\label{fig:3} Simulated OFC generation in a singly resonant SHG resonator. Parameters are similar to the ones used in Ref.~\cite{Leo2016a}. (a) Fundamental comb (orange) power (excluding the power of pump frequency component) and second-harmonic field (blue) power as a function of pump power for the detuning $\Delta_A t_{{\rm R},A}=0$. Fundamental combs generated with parameters (b) $P_{in}=170$ mW, $\Delta_A t_{{\rm R},A}=0.004$; (c) $P_{in}=2000$ mW, $\Delta_A t_{{\rm R},A}=-0.02$; and (d) $P_{in}=2000$ mW, $\Delta_A t_{{\rm R},A}=-0.01$. (e)-(g) Corresponding second-harmonic combs. The insets in (b)-(g) are corresponding temporal patterns.}
\end{figure*}

The SHG cavity containing a quadratically nonlinear crystal with length $L=15$ mm is singly resonant for the fundamental field, namely, $\theta_B=1$. The wavelength of the pump laser is 1064 nm. By the definition of the amplitude $A$ in Eq.~(\ref{eq:4}), the intensity $I$ is given by $I=2n\epsilon_0c|A|^2$. In our simulation, the amplitude of the pump laser electric field is obtained by $I_{in}=P_{in}/S$, where $P_{in}$ is the pump power, $I_{in}$ is the intensity of the pump laser, and $S$ is the effective area. The typical value of $\chi^{(2)}$ is 5 pm/V. The effective area $S=51.969$ ${\rm \mu m}^2$ is calculated from the nonlinear coupling coefficient $\kappa$ in Ref.~\cite{Leo2016a}. From the Sellmeier equation for the 5\% MgO doped ${\rm LiNbO}_3$ \cite{Gayer2008} at a crystal temperature $T_0=39.5$ $^{\circ}$C, the refractive indices $n_A=2.152$ and $n_B=2.229$ are obtained. Because the fundamental field and the second-harmonic field are both restricted to their center frequencies, the approximation $\omega_{A,\mu}\approx\omega_0$ and $\omega_{B,\xi}\approx2\omega_0$ are used for simplicity. The other parameters are set as the following: $\theta_A=0.02$, $\alpha_{iA}=\alpha_{iB}=1.309$ m$^{-1}$, the FSR of fundamental components $D_{1A}/(2\pi)=9.053$ GHz, the FSR of second-harmonic components $D_{1B}/(2\pi)=8.174$ GHz, and the second-order dispersions $D_{2A}=-0.1055$ MHz and $D_{2B}=-0.2533$ MHz. The second-harmonic detuning $\Delta_B$ is set to 0 for simplicity and the pump detuning $\Delta_A$ remains as a free parameter.

The simulated power of the second-harmonic and intracavity fundamental combs are plotted in Fig.~\ref{fig:3}(a) as a function of the input pump laser power for the detuning $\Delta_A t_{{\rm R}}=0$. The pump frequency component is not included in the plot. Fig.~\ref{fig:3}(a) shows that when the pump power is below the threshold for the emergence of the frequency comb, the second-harmonic power grows almost linearly with the pump power. Once the pump power reaches the threshold, however, the power of second-harmonic is saturated. It is because the second-harmonic power is converted back to the fundamental components. The simulated frequency combs with different pump powers and detunings are also plotted in Figs.~\ref{fig:3}(b)-(g). The periodic temporal profiles in Figs.~\ref{fig:3}(b) and (c) indicate the combs are coherent. However, the comb in Fig~\ref{fig:3}(d) is plagued by noise. The stability analysis of Kerr OFCs has shown that for different detunings and pump powers, there are a lot of comb states, such as Turing patterns, chaos and solitons \cite{Godey2014}. Fig~\ref{fig:3}(b)-(g) show that different comb states of quadratic OFCs for different detunings and pump powers may also exist. Our simulated results are in agreement with the results from the time-domain model in Ref.~\cite{Leo2016a}.

To demonstrate our model is also valid for third-order nonlinearity, we present the second example as the OFC generation in a resonator containing Kerr material. The third-order nonlinearity in such materials is dominant. The nonlinear polarization can be written as
\begin{equation}
{\rm P}_\mu^{{\rm NL}}={\rm P}_\mu^{(3)}=p_\mu^{(3)}e^{-i\omega_\mu t}+{\rm c.c.}.
\label{eq:19}
\end{equation}
Here the interaction of all frequency components are considered as
\begin{eqnarray}
p_\mu^{(3)}=&&\epsilon_0\Bigg[\sum_{\alpha_1,\beta_1,\gamma_1}\chi_{\alpha_1\beta_1\gamma_1}^{(3)}{\cal E}_{\alpha_1}{\cal E}_{\beta_1}{\cal E}_{\gamma_1}\nonumber\\
&&\times\delta\left(\omega_{\alpha_1}+\omega_{\beta_1}+\omega_{\gamma_1}-\omega_{\mu}\right)\nonumber\\
&&+3\sum_{\alpha_2,\beta_2,\gamma_2}\chi_{\alpha_2\beta_2\gamma_2}^{(3)}{\cal E}_{\alpha_2}{\cal E}_{\beta_2}^*{\cal E}_{\gamma_2}\nonumber\\
&&\times\delta\left(\omega_{\alpha_2}-\omega_{\beta_2}+\omega_{\gamma_2}-\omega_{\mu}\right)\nonumber\\
&&+3\sum_{\alpha_3,\beta_3,\gamma_3}\chi_{\alpha_3\beta_3\gamma_3}^{(3)}{\cal E}_{\alpha_3}{\cal E}_{\beta_3}^*{\cal E}_{\gamma_3}^*\nonumber\\
&&\times\delta\left(\omega_{\alpha_3}-\omega_{\beta_3}-\omega_{\gamma_3}-\omega_{\mu}\right)\Bigg],
\label{eq:20}
\end{eqnarray}
where $\chi^{(3)}$ is the third-order nonlinear optical susceptibility, and $\alpha_j$, $\beta_j$ and $\gamma_j$ with $j=1,2,3$ denote different frequency components. The first term in Eq.~(\ref{eq:20}) describes the sum-frequency generation processes, including the third-harmonic generation as a special case. The second term describes self-phase modulation and cross-phase modulation. The third term describes the inverse process of the first term. Similar to the second term in Eq.~(\ref{eq:6}), there is a factor 3 in the second and third terms here due to the permutation of the indices. The first term and third term require a (quasi-) phase-match condition. However, for a compact platform, such as the SiN microring resonator, the phase-match condition is usually not satisfied. Therefore, only the second term in Eq.~(\ref{eq:20}) is considered. To compare with the time-domain model LLE for Kerr OFCs \cite{Coen2013}, here the optical field is restricted to its center frequency so the higher-order dispersion, the phase mismatch $\Delta k$, and the frequency dependence of parameters can be neglected. Eqs.~(\ref{eq:8}) and (\ref{eq:9}) become
\begin{equation}
\frac{dA_\mu}{dz}=-\left(\frac{\alpha_{i}}{2}+i\delta_\mu\right)A_\mu+\frac{i\omega_\mu}{2n_0c}{\cal N},
\label{eq:21}
\end{equation}
\begin{equation}
A_\mu^{(m+1)}(0)=\sqrt{1-\theta}A_\mu^{(m)}(L)+\sqrt{\theta}A_{in}\delta(0),
\label{eq:22}
\end{equation}
where ${\cal N}=3\sum_{\alpha,\beta,\gamma}\chi^{(3)}A_\alpha A_\beta^* A_\gamma\delta(\alpha-\beta+\gamma-\mu)$. Using Eq.~(\ref{eq:10}), $\delta_\mu$ is written as
\begin{equation}
\delta_\mu=\left(\Delta+\frac{1}{2}D_2\mu^2\right)t_{{\rm R}}/L,
\label{eq:23}
\end{equation}
where $\Delta=\omega_{resonant,\omega_0}-\omega_0$ is defined to represent the pump detuning.

In the simulation, the pump wavelength is 1562 nm. Other parameters are set as the following: the coupling ratio $\theta=0.009$, the linear loss $\alpha_i=14.32$ m$^{-1}$, the refractive index $n_0=2.112$, and the third-order nonlinear optical susceptibility $\chi^{(3)}=10^{-21}$ ${\rm m}^2/{\rm V}^2$. Because the optical field is restricted to its center frequency, the approximation $\omega_\mu=\omega_0$  is used for simplicity. Again, we use the formula $I_{in}=P_{in}/S$ with $P_{in}=575$ mW to obtain the amplitude of the pump laser electric field. The effective area $S=0.255$ ${\rm \mu m}^2$ is very small because of the size of a microring resonator, and it corresponds to the nonlinearity coefficient $\gamma=n_2\omega_0/(c S)=1$ ${\rm W}^{-1}{\rm m}^{-1}$ used in the time-domain model LEE with $n_2$ being the nonlinear refractive index. The radius of the microring is 100 ${\rm \mu m}$ and hence the length $L=628.139$ ${\rm \mu m}$. The microring has the FSR $D_1/(2\pi)=226$ GHz and the second-order dispersion $D_2=13.887$ MHz which corresponds to the dispersion coefficient $\beta_2=-48.5$ ${\rm ps}^2/{\rm km}$ used in LLE.

Due to the double balance between loss and parametric gain and between dispersion and nonlinearity, there is a special state in time domain called single dissipative Kerr soliton state which corresponds to a fully coherent and low noise Kerr OFC. Scanning the detuning from blue to red detuning is the most common way to enter a single soliton state \cite{Herr2014a,Guo2017}. When a single soliton is obtained through the detuning sweeping scheme, the precise detuning is sensitive to the random seed used as the initial conditions. In the simulation, once a single soliton is formed, the detuning is set back to $\Delta t_{\rm R}=0.045$ so the simulated results obtained from different models can be compared under the identical parameters.
\begin{figure}[b]
\includegraphics[width=240pt]{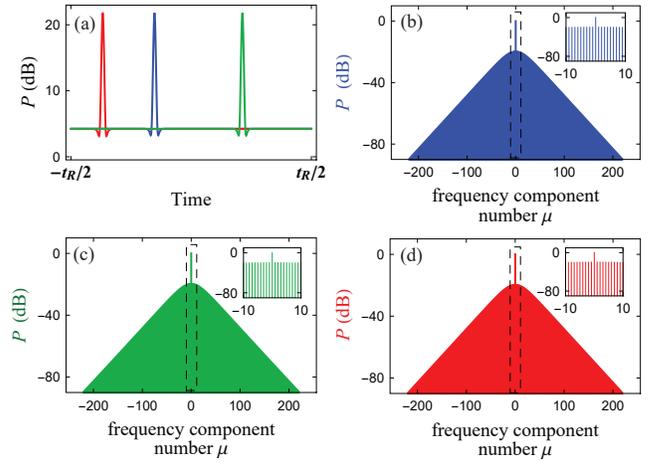}
\caption{\label{fig:4} Solitons and Kerr OFCs simulated by full map equations (blue) and LLE (green) as well as mean-field equations (red). (a) Solitons. (b)-(d) Kerr OFCs. The insets in (b)-(d) are zoom-in spectra. Results obtained by our frequency-domain models and LLE are identical.}
\end{figure}

Fig.~\ref{fig:4}(a) shows solitons simulated by using the full map model (blue) and LLE (green). For a small coupling ratio $\theta$, the mean-field equations in Eq.~(\ref{eq:12}) are also valid. The soliton obtained by using the mean-field equations is plotted as the red line in Fig.~\ref{fig:4}(a). The position of the soliton in the cavity depends on the random initial condition and it does not affect the comb spectra. The corresponding OFC spectra are shown in Fig.~\ref{fig:4}(b)-(d). These combs have smooth envelopes. If the dispersion is perturbed by the avoided mode crossing effect at a specific frequency position, combs will have the characteristic up-down feature as reported in Ref.~\cite{Herr2014}. Because here we truncate the cavity dispersion at second-order, these combs have symmetrical structures. If the dispersion is near the ZDP, the third-order and even the higher-order dispersion must be considered. In such cases, combs may be extended by stimulating Cherenkov radiation \cite{Bras2016}. These simulated results obtained by using different models are in an excellent agreement.

\section{\label{sec:level4}Conclusion}
In conclusion, we have developed a general framework in the frequency domain for simulating OFC generation in resonators that contain second- and third-order nonlinearities. The frequency dependence of parameters can be readily accounted for in Eqs.~(\ref{eq:8}) and (\ref{eq:9}). Particularly, we demonstrated that the cavity dispersion could be expressed accurately or truncated beyond a certain order by expanding the resonant frequencies in a Taylor series. We presented two examples to validate the model. The first example is OFCs generated via the quasi-phase-matched cavity-enhanced second-harmonic generation. The simulated results are identical to the ones obtained in the time-domain model \cite{Leo2016a}. The stability analysis of quadratic OFCs revealed various comb states, similar to the ones in Kerr OFCs \cite{Godey2014}. The second example is Kerr OFCs in a micro-resonator. The simulated results obtained from full map equations and LLE as well as mean-field equations are in an excellent agreement. Recently, OFCs generated in resonators containing both second- and third-order nonlinearities have been reported \cite{Xue2017,Hao2020}. This work can contribute to understanding OFC generation in various resonators, especially when multi-order nonlinearities are present and the frequency dependence of system parameters is not negligible.


\providecommand{\noopsort}[1]{}\providecommand{\singleletter}[1]{#1}%

\end{document}